# Tunning the number of chiral edge channels in a fixed quantum anomalous Hall system

Peng Deng[1,2†*], Yulei Han[3†], Peng Zhang[2], Su Kong Chong[2], Zhenhua Qiao[4,5*], and Kang L. Wang[2*]

[1]*Beijing Academy of Quantum Information Sciences, Beijing 100193, China*

[2]*Department of Electrical and Computer Engineering, University of California, Los Angeles, California 90095, United States*

[3]*Department of Physics, Fuzhou University, Fuzhou, Fujian 350108, China*

[4]*International Center for Quantum Design of Functional Materials, CAS Key Laboratory of Strongly-Coupled Quantum Matter Physics, and Department of Physics, University of Science and Technology of China, Anhui 230026, China*

[5]*Hefei National Laboratory, University of Science and Technology of China, Hefei 230088, China*

[†]These authors contributed equally: Peng Deng, Yulei Han

[*]Corresponding authors: dengpeng@baqis.ac.cn, qiao@ustc.edu.cn, wang@ee.ucla.edu

**Abstract:**

Quantum anomalous Hall (QAH) insulators exhibit chiral edge channels characterized by vanishing longitudinal conductance and quantized Hall conductance of $Ce^2/h$, wherein the Chern number $C$ is an integer equal to the number of the parallel chiral edge channels. These chiral edge channels conduct dissipationless transport in QAH insulators, making them pivotal for applications in low-consumption electronics and topological quantum computing. While the QAH effect with multiple chiral edge channels (*i.e.*, $C >1$) has been demonstrated in multilayers consisting of magnetic topological insulators and normal insulators, the channel number remains fixed for a given sample. Here, we unveil the tunability of the number of chiral edge channels within a single QAH insulator device. By tuning the magnetization of individual layers within the multilayer system, Chern insulating states with different Chern numbers are unveiled. The tunable Chern number was corroborated by our theoretical calculations. Furthermore, we conducted layer-dependent calculations to elucidate the contribution of the Chern number from different layers in the multilayer. Our findings demonstrate an extra degree of freedom in manipulating the chiral edge channels in QAH insulators. This newfound tunability offers extra dimension for the implementation of the QAH-based multi-channel dissipationless transport.

**I. Introduction**

The quantum anomalous Hall (QAH) effect, the counterpart of the quantum Hall effect without an external magnetic field, features an insulating bulk and chiral edge channels propagating at the boundaries [1-7]. At the zero magnetic field, the Hall conductance of a QAH insulator exhibits a precise quantization of $Ce^2/h$, wherein $e$ is the elementary charge, $h$ is the Planck constant, and the integer $C$ is known as the Chern number [8]. The Chern number characterizes the nontrivial topology of the bulk band structure, equating to the number of chiral edge channels in the system. In quantum Hall materials, the Chern number can be experimentally tuned by varying the magnetic field or applying a gate voltage. In QAH systems, however, such tunability is less demonstrated.

The QAH effect has been observed in several material systems including Cr-doped (Bi,

Sb)$_2$Te$_3$ (CBST) [9-11], V-doped (Bi, Sb)$_2$Te$_3$ (VBST) [12], twisted bilayer graphene [13], MnBi$_2$Te$_4$ [14-16], and WSe$_2$/MoTe$_2$ bilayer [17], typically exhibiting a prevalent Chern number of 1. While the discovery of the $C = 1$ QAH effect has spurred significant research, many efforts have been devoted to finding high-Chern number QAH insulators in which multiple chiral edge channels could significantly lower the energy consumption and enhance device performance [18-23]. Notably, QAH effect with $C = 2$ has been reported in MnBi$_2$Te$_4$ flakes [16] and twisted monolayer-bilayer graphene [24]. In a parallel route, the high-Chern number QAH effect has also been realized in molecular beam epitaxy (MBE) grown multilayers comprising alternating stacking of $C = 1$ QAH insulators and normal insulators. In the Cr, V-codoped BST/CdSe multilayers [25] and CBST/BST multilayers [26], high-Chern number QAH effects up to $C = 4$ and $C = 5$ have been reported, respectively. In these multilayers, however, the value of $|C|$ remains fixed for a specific sample. To tune the Chern number, the composition of the material must be altered. This greatly limits the practical application of the high-Chern number QAH effect in electronic devices.

Here, we report the experimental realization of the high-Chern number QAH effect in a QAH multilayer consisting of VBST, BST, and CBST layers. By controlling the magnetization of the CBST and VBST layers using an external magnetic field, Chern insulating states with different Chern numbers of $C = 0, \pm 1$, and $\pm 2$ are revealed in a single sample. The tunable Chern number was further elucidated by our theoretical calculation, showing the phase diagram of the QAH multilayer. Moreover, layer-dependent calculations were carried out to understand the contribution of the Chern number from each layer in the multilayer system.

## II. Constructing different types of QAH multilayers

The QAH multilayers are grown by MBE on semi-insulating GaAs(111)B substrates, and the details of the growth can be found in Supplementary Materials. Using BST, CBST, and VBST layers as elementary building blocks and assembling them like Lego bricks enables the construction of multilayers with diverse structures, resulting in QAH insulators with distinct properties. Figure 1(a) schematically depicts a 3-quintuple (QL)-CBST/5-QL-BST/3-QL-CBST multilayer. The field dependence of the longitudinal and Hall conductance, $\sigma_{xx}$ and $\sigma_{xy}$, of the multilayer are presented in Fig. 1(b) and 1(c), respectively. When fully magnetized, the multilayer exhibits vanishing longitudinal conductance and quantized Hall conductance of $\sigma_{xy} = \pm e^2/h$, representing QAH states of $C = \pm 1$. Figure 1(d) is the flow diagram of the $\sigma_{xy}$ vs. $\sigma_{xx}$ plot for the 3-QL-CBST/5-QL-BST/3-QL-CBST multilayer. In the plot, it is evident that the $C = 1$ state transitions directly into the $C = -1$ state, and no sign of the $C = 0$ state is observed during the transition. The absence of the $C = 0$ state conveys two crucial pieces of information. Firstly, the 3-QL-CBST/5-QL-BST/3-QL-CBST sample, with a total thickness of 11 QL, is sufficiently thick to prevent the formation of the normal insulator state due to the hybridization between the top and bottom surface states [11,27]. Secondly, the magnetization in the top and bottom CBST layers flips simultaneously during the reversal process. Otherwise, a $C = 0$ plateau signaling the axion insulator would be observed when the top and bottom magnetization points to opposite directions, as in the case of the VBST/BST/CBST multilayer [28-35].

By replacing the top CBST layer with a VBST layer, we construct a 3-QL-VBST/5-QL-BST/3-QL-CBST multilayer [Fig. 1(e)]. Its field dependence of $\sigma_{xx}$ and $\sigma_{xy}$ are presented in Figs. 1(f) and 1(g), respectively. In this VBST/BST/CBST multilayer, the observed vanishing longitudinal conductance and quantized Hall conductance signify the existence of the $C = \pm 1$ QAH states when the system is fully magnetized. Moreover, since the CBST layer has a smaller magnetic coercive field than the VBST layer (see Fig. S2 in Supplementary Materials), the magnetization in the top and bottom layers flips independently during the field sweep, and the axion insulating state forms during the magnetization reversal process. Indeed, zero Hall plateaus are observed in the field range between the coercive fields of CBST and VBST layers [Fig. 1(g)]. The existence of the $C = 0$ is also evidenced in the $\sigma_{xy}$ vs. $\sigma_{xx}$ plot, as shown in Fig. 1(h). In the VBST/BST/CBST multilayer, the system experiences a $C = 0$ state when transitioning between $C = -1$ and $C = 1$ states, in sharp contrast to the case of the CBST/BST/CBST multilayer.

By increasing the repeat of the superlattice in the multilayer, the high-Chern number QAH effect can be realized [25,26]. To this end, we grow a 3-QL-VBST/5-QL-BST/3-QL-CBST/5-QL-BST/3-QL-CBST multilayer [Fig. 1(i)]. Figures 1(j) and 1(k) present the field dependence of $\sigma_{xx}$ and $\sigma_{xy}$ of the VBST/BST/CBST/BST/CBST multilayer, respectively. As can be seen, the Hall conductance exhibits quantized plateaus of $\sigma_{xy} = \pm e^2/h$ and $\pm 2e^2/h$, with concomitant local minima in $\sigma_{xx}$, indicating the presence of the Chern insulating states with $C = \pm 1$ and $\pm 2$. These distinct states correspond to different magnetization configurations in the multilayer: the system is in the $C = \pm 2$ states when magnetization in the CBST and VBST layers is parallelly aligned, and is in the $C = \pm 1$ states when magnetization is antiparallelly aligned. Notably, in the VBST/BST/CBST/BST/CBST multilayer, the magnetization reversal process in the middle and bottom CBST layers occurs simultaneously, as evidenced by the minor loop scans shown in Fig. S3. Therefore, these types of configurations are only allowed stable ones in the multilayer.

Besides the local minimums of $\sigma_{xx}$ corresponding to the $C = \pm 1$ and $\pm 2$ states, an additional pair of local minimums emerges precisely when $\sigma_{xy} = 0$, as can be better resolved in the zoomed-in plots presented in the right panels of Figs. 1(j) and 1(k). This $C = 0$ state does not develop into a plateau, as in the VBST/BST/CBST case; instead, it only manifests as a kink in the $\sigma_{xy}$ curves. This is because the $C = 0$ state occurs at the point somewhere in the middle of the magnetization reversal process of the CBST layer. In this case, the out-of-plane component of the magnetization in the middle and bottom layers combined cancels out with the magnetization in the VBST layer, giving a net zero magnetization in the system. However, such a magnetic configuration is unstable, leading to the transient $C = 0$ state. Despite being the case, all the $C = 0, \pm 1$, and $\pm 2$ states are clearly visible in the $\sigma_{xy}$ vs. $\sigma_{xx}$ plot, as shown in Fig. 1(i). We would like to note that in comparison to the case of CBST/BST/CBST or VBST/BST/CBST multilayer, the VBST/BST/CBST/BST/CBST multilayer has a larger residual longitudinal conductance. In particular, the $C = 0$ state has an even greater residual longitudinal conductance, suggesting that the measuring temperature may not be sufficiently low to eliminate the thermally activated carriers. More importantly, it indicates that the Fermi level in the VBST/BST/CBST/BST/CBST does not reside as deeply within the exchange gap as in the CBST/BST/CBST or VBST/BST/CBST, for the reasons discussed below.

**III. Temperature dependence**

Figures 2(a) and 2(b) present the field dependence of the $\sigma_{xx}$ and $\sigma_{xy}$ under different temperatures for the VBST/BST/CBST/BST/CBST multilayer, and Figs. 2(c) and 2(d) are the zoomed-in plot around the $C = 0$ state. For clarity, only data taken under leftward field sweep are displayed (the rightward field sweep data are shown in Fig. S4). The temperature evolutions of the $\sigma_{xx}$ and $\sigma_{xy}$ for each Chern insulator state are summarized in Figs. 2(e) and 2(f), respectively. Upon decreasing temperature, $\sigma_{xy}$ approaches the corresponding quantized value for all states, meanwhile, all $\sigma_{xx}$ displays the tendency towards approaching zero.

The value of the exchange gap Δ for each Chern insulating state can be obtained by fitting the corresponding temperature-dependent $\sigma_{xx}$ with the Arrhenius relationship $\sigma_{xx} \sim e^{-\Delta/k_B T}$, wherein $k_B$ is the Boltzmann constant [Fig. S5]. The fitted gap values are summarized in Table 1. The exchange gap values obtained in the CBST/BST/CBST and VBST/BST/CBST multilayers are comparable to those reported for uniformly doped CBST [36], while the gap values for Chern insulating states in the VBST/BST/CBST/BST/CBST multilayer are significantly smaller, consistent with the observation of larger residual longitudinal conductance in this multilayer. The reduction of the exchange gap in the VBST/BST/CBST/BST/CBST multilayer is likely caused by the existence of the metallic states on the side surfaces and the local potential fluctuations resulting from disorder, both of which become more pronounced in the thicker sample. Moreover, for the $C = \pm 1$ and $\pm 2$ states, both the CBST and VBST layers exhibit a single magnetic domain, whereas, for the $C = 0$ case, the CBST layers are in a multi-domain state. This additional disorder further reduces the gap value of the $C = 0$ state.

The observed $C = 0$ phase in the VBST/BST/CBST/BST/CBST multilayer strongly supports the existence of the axion insulating state. Firstly, as demonstrated above, both $\sigma_{xx}$ and $\sigma_{xy}$ approach zero upon decreasing temperature, a characteristic feature of the axion insulating state. Secondly, the $C = 0$ states emerge precisely when the net magnetization of the multilayer is zero, aligning with the criteria for the axion insulating state. Thirdly, the observed $C = 0$ state is not indicative of a normal insulator state. As discussed above, in the CBST/BST/CBST multilayer with a total thickness of 11 QL, the hybridization between the top and bottom surface states is insufficient to form a normal insulator. In the 19 QL thick VBST/BST/CBST/BST/CBST multilayer, the hybridization is even smaller.

## IV. Phase diagram of the QAH multilayer

To better understand the underlying mechanism for the existence of different Chern insulating phases in the QAH multilayer, we construct a tight-binding lattice model from the continuum model for topological insulators [37]:

$$H = \sum_i (E_0 + m_{exc}) C_i^+ C_i + \sum_{\alpha,i} C_{i+1}^+ T_\alpha C_i + + h.c.$$

where $E_0 = (M_0(l) - 6B)\, \sigma_z \otimes s_0$ and $T_\alpha = B\sigma_z \otimes s_0 + \frac{iA}{2}\sigma_x \otimes s_\alpha$ with $\alpha = x, y, z$, $m_{exc} = m(l)\sigma_0 \otimes s_z$. The Pauli matrices $\sigma$ and $s$ represent the orbital and spin degrees of freedom, respectively. The multilayer configuration of the VBST/BST/CBST/BST/CBST is simulated by the layer-dependent terms $M_0(l)$ and $m(l)$,

$$M_0(l), m(l) = \begin{cases} M_0, 0 & for\ BST\ layer \\ M_{CBST}, g_{CBST} & for\ CBST\ layer \\ M_{VBST}, g_{VBST} & for\ VBST\ layer \end{cases}$$

wherein $m(l)$ is the magnetization and $M_0(l)$ determines the topological trivialness of the layer in the infinite thickness limit, i.e., $M_0(l) < 0$ for trivial phase and $0 < M_0(l) < 4B$ for nontrivial phase. In our calculation, we set $A = 1.0$, $B = 0.6$, and $M_0 = 1.0$.

Figure 3 presents the phase diagram of the 3-QL-CBST/5-QL-BST/3-QL-CBST/5-QL-BST/3-QL-VBST multilayer with varying magnetization in CBST and VBST layers. Without loss of generality, the values of $M_{CBST}$ and $M_{VBST}$ are set to be 0.2, 0, and -0.2. As shown in the plot, the system transitions between different Chern insulating states when $g_{CBST}$ or $g_{VBST}$ is being tuned. Note that the result of the phase diagram does not qualitatively depend on the sign of $M_{CBST/VBST}$. This interesting observation in calculation is consistent with the fact that the high-Chern number QAH effect is realized both in distinctly different multilayers reported in this study and Ref. 26. In our work, the CBST and VBST layers are grown using the same recipe to grow the $C = 1$ QAH insulators [38,39], corresponding to the $M_{CBST/VBST} > 0$ case; the CBST/BST/CBST/BST/CBST multilayer reported in Ref. 26, however, is heavily doped and topological trivial, i.e., $M_{CBST/VBST} <$

0. In both cases, the high-Chern number QAH effect is realized.

## V. Layer-resolved calculation

To get the distribution of $C$ from each layer in the VBST/BST/CBST/BST/CBST multilayer, layer-resolved calculations were performed for the 3-QL-CBST/5-QL-BST/3-QL-CBST/5-QL-BST/3-QL-VBST multilayer. The partial Chern number $C_z(l)$ is the projection of the Chern number onto the $l$-th layer, and is written as [40]:

$$C_z(l) = \frac{-4\pi}{A} \operatorname{Im} \frac{1}{N_k} \Sigma_k \Sigma_{vv'c} X_{vck} Y^+_{v'ck} \rho_{v'vk}(l),$$

where $X$ ($Y$) is the position operator along the $x$ ($y$) direction, and $\rho_{v'vk}(l) = \Sigma_{j \in l} \psi^*_{vk}(j) \psi_{v'k}(j)$ is the Bloch representation of the projection onto layer $l$. The sum of $C_z(l)$ from the first to the $l$-th layer is denoted as $C_{int}(l)$, it is the Chern number accumulated from the bottom layer to the layer of interest.

Figures 4(a)-4(e) present the layer-dependent $C_z(l)$ and $C_{int}(l)$ for $C$ = 0, ± 1, and ± 2 states when $M_{CBST/VBST}$ = 0.2 (the results for $M_{CBST/VBST}$ = -0.2 are shown in Fig. S8). Notably, the partial Chern number is predominately contributed by magnetic layers, with minimum the contribution from nonmagnetic layers. Consequently, the $C_{int}$ (= 0.5 or 1.5) plateaus are evident in the nonmagnetic layers. Figures 4(f)-4(j) summarize the $C_z(l)$ contributed from BST, CBST, and VBST blocks. In $C$ = ± 1 and ± 2 states, where the magnetic layers are in the single domain state, the partial Chern number in the middle CBST block is twice as large as in the top VBST and bottom CBST blocks. This is attributed to the fact that the middle CBST layer interfaces with two BST layers, while the top VBST or bottom CBST layer interfaces with only one. The $C$ = 0 state, however, reveals a very distinct distribution of $C_z(l)$: the partial Chern number is mostly contributed from the top and bottom blocks with opposite signs, giving rise to a net zero Chern number. This calculated result supports the argument that the $C$ = 0 state in the VBST/BST/CBST/BST/CBST multilayer is an axion insulating state. Finally, comparing the results in these states reveals a localized impact of magnetization on $C_z(l)$ distribution, limited to the corresponding magnetic blocks and their neighboring blocks. For instance, in the $C$ = 1 [Fig. 4(d)] and $C$ = 2 states [Fig. 4(e)], where the magnetization in CBST layers is the same while that in VBST is in opposite directions, the disparity in $C_z(l)$ distribution between these two scenarios is noticeable only in the VBST layer and the interfacing 1 QL BST.

## VI. Conclusions

In conclusion, our work reveals the presence of Chern insulating states with tunable Chern numbers in MBE grown QAH multilayers composed of BST, CBST, and VBST layers. The manipulation of the magnetization in magnetic layers allows for the tuning of the number of the chiral edge states in the multilayer. Model calculations were conducted to understand the existence of these Chern insulating states, and a phase diagram of the QAH multilayer is unveiled. Our layer-revolved calculations further elucidate the contributions of the Chern number from different layers within the multilayer. In the future, more Chern insulating states with higher Chern numbers can potentially be realized in QAH multilayers with more complex structures, e.g., multilayers consisting of CBST, VBST, and Cr, V co-doped BST layers. Our work demonstrated the QAH multilayer as a playground to explore the rich turnability of the chiral edge channels.

**Figures and figure captions**

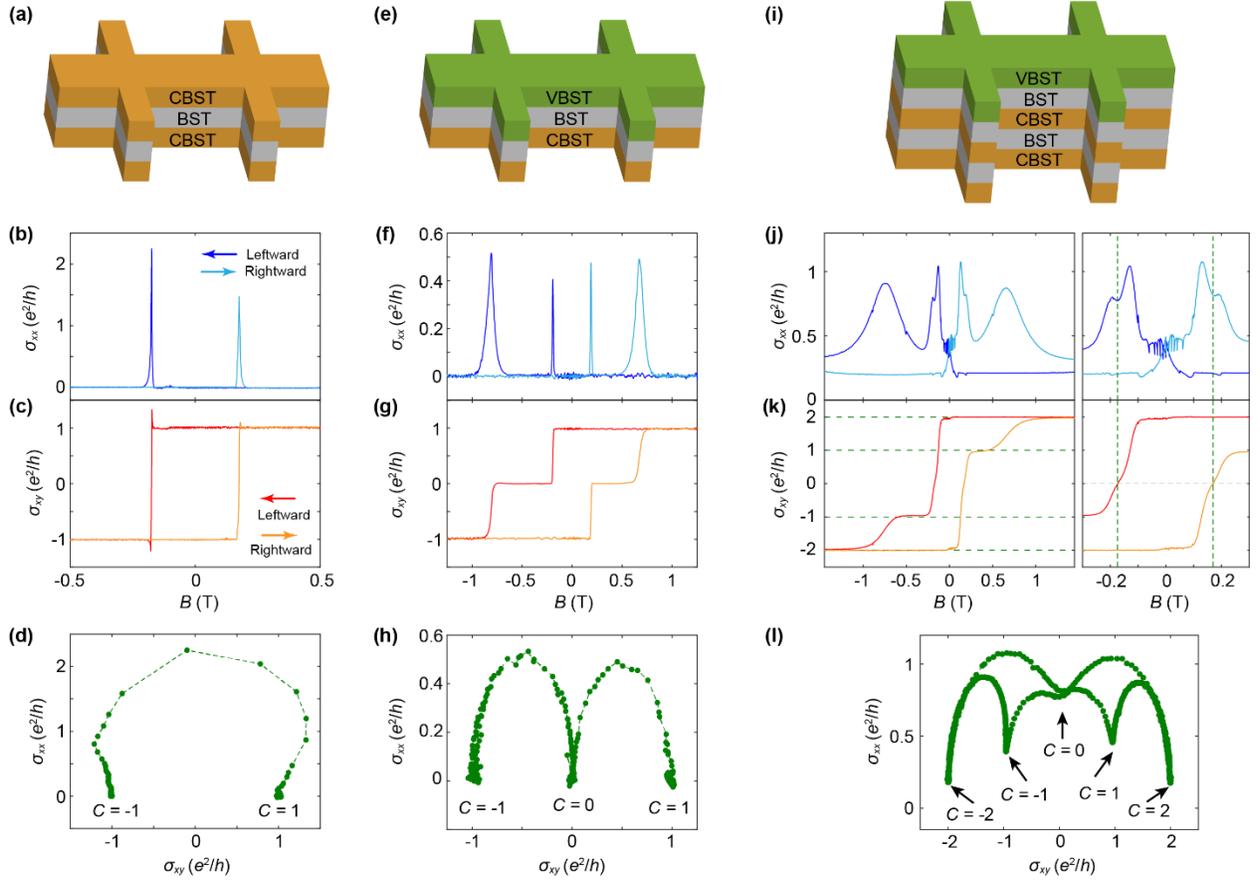

FIG. 1. (a) Schematic layout of a 3-QL-CBST/5-QL-BST/3-QL-CBST multilayer. (b, c) Field dependence of $\sigma_{xx}$ and $\sigma_{xy}$ of the CBST/BST/CBST multilayer measured at 100 mK. (d) Flow diagram of ($\sigma_{xy}$, $\sigma_{xx}$) of the CBST/BST/CBST multilayer. (e) Schematic layout of a 3-QL-VBST/5-QL-BST/3-QL-CBST multilayer. (f, g) Field dependence of $\sigma_{xx}$ and $\sigma_{xy}$ of the VBST/BST/CBST multilayer measured at 100 mK. (h) Flow diagram of ($\sigma_{xy}$, $\sigma_{xx}$) of the VBST/BST/CBST multilayer. (i) Schematic layout of a 3-QL-VBST/5-QL-BST/3-QL-CBST/5-QL-BST/3-QL-CBST multilayer. (i, k) Field dependence of $\sigma_{xx}$ and $\sigma_{xy}$ of the VBST/BST/CBST/BST/CBST multilayer measured at 50 mK. Panels on the right are the zoomed in plots around the $C = 0$ states. (l) Flow diagram of ($\sigma_{xy}$, $\sigma_{xx}$) of the VBST/BST/CBST/BST/CBST multilayer.

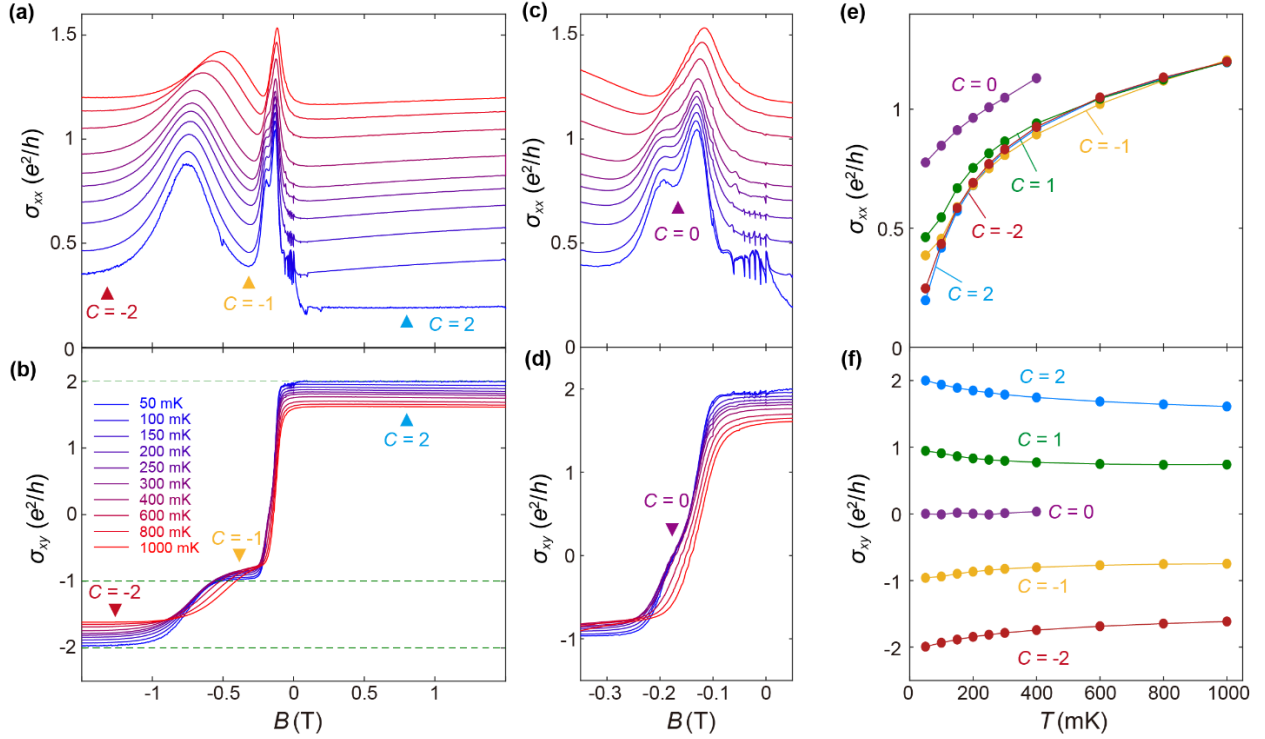

FIG. 2. (a, b) Field dependence of $\sigma_{xx}$ and $\sigma_{xy}$ of the 3-QL-VBST/5-QL-BST/3-QL-CBST/5-QL-BST/3-QL-CBST multilayer under different temperatures. Only leftward field scan result is shown for clarity. (c, d) Zoomed in plots of field dependence of $\sigma_{xx}$ and $\sigma_{xy}$ for the $C = 0$ state. (e, f) Temperature evolution of the $\sigma_{xx}$ and $\sigma_{xy}$ for the $C = 0, \pm 1$, and $\pm 2$ states.

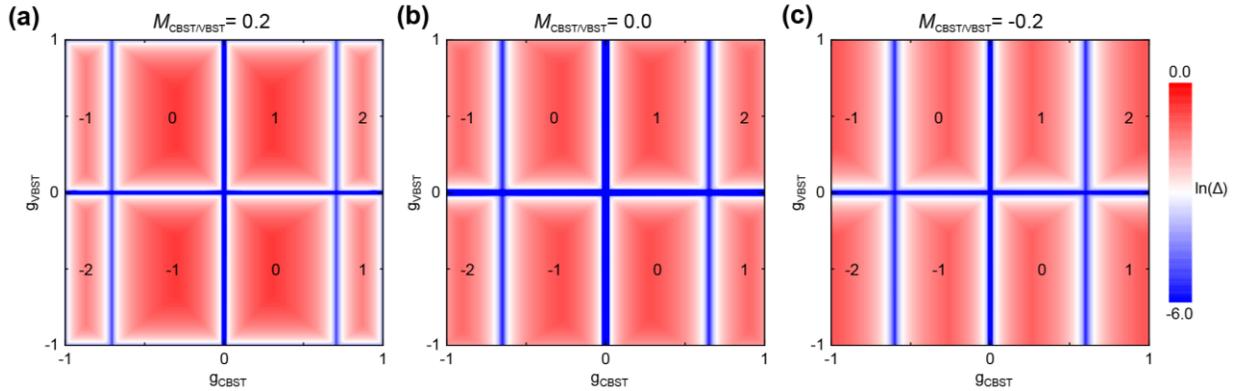

FIG. 3. Topological phase diagram of VBST/BST/CBST/BST/CBST multilayer in ($g_{CBST}$, $g_{VBST}$) plane for (a) $M_{CBST/VBST} = 0.2$, (b) $M_{CBST/VBST} = 0$, and (c) $M_{CBST/VBST} = -0.2$.

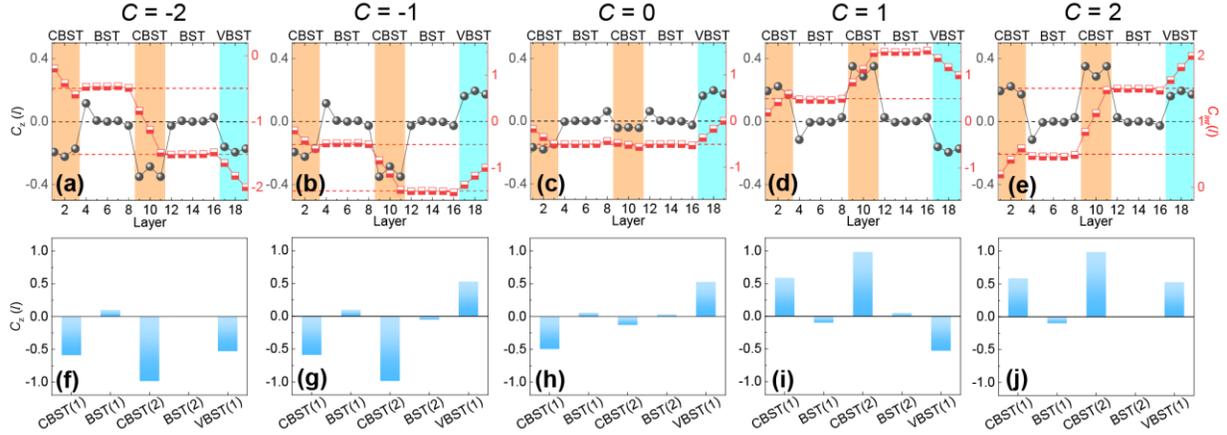

FIG. 4. (a)-(e) Layer-resolved partial Chern number $C_z(l)$ (grey) and its integral $C_{int}(l)$ (red) as a function of layer number $l$ in the VBST/BST/CBST/BST/CBST multilayer system for different Chern insulating states. The orange, white and cyan background colors represent the layers of CBST, BST and VBST, respectively. (f)-(j) The corresponding integral of $C(l)$ in each CBST, BST, or VBST block for the Chern insulating states.

Table 1. Exchange gap Δ.

| Chern insulating states | $C$ = 2 | $C$ = 1 | $C$ = 0 | $C$ = -1 | $C$ = -2 | CBST/BST/CBST ($C$ = 1) | VBST/BST/CBST ($C$ = 1) |
|---|---|---|---|---|---|---|---|
| $\Delta/k_B$ (mK) | 126.7 | 94.4 | 37.7 | 120.1 | 122.8 | 1012.9 | 1415.2 |

**Acknowledgments**

This work was supported by National Natural Science Foundation of China (No. 12304189) and Beijing Natural Science Foundation (No. 1232035). Work at UCLA was supported by the NSF (No. 1936383 and No. 2040737) and the U.S. Army Research Office MURI program (No.

W911NF-20-2-0166). Work at USTC was supported by the National Natural Science Foundation of China (No. 11974327, No. 12004369) and Anhui Initiative in Quantum Information Technologies (No. AHY170000). Work at FZU was supported by Natural Science Foundation of Fujian Province (2022J05019). The supercomputing service of USTC and AM-HPC is gratefully acknowledged.